\documentclass[aps,preprint,superscriptaddress,longbibliography]{revtex4-1}

\usepackage{amsmath,amssymb}
\usepackage{graphicx}
\usepackage{epstopdf}
\usepackage{bm}
\usepackage{color}

\newcommand{\fese}     {FeSe}
\newcommand{\se}	{$^{77}$Se}
\newcommand{\slr} 	{$T_1^{-1}$}
\newcommand{\slrt} 	{$(T_1T)^{-1}$}

\newcommand{\chispin} 	{$\chi_\text{spin}$}

\usepackage{multibib}
\newcites{S}{Supplementary References}

\newcommand{\bc}[1]{\textbf{\sffamily #1}}

\begin{document}

\title{Orbital-driven nematicity in FeSe}

\author{S.-H. Baek}
\email[]{sbaek.fu@gmail.com}
\affiliation{IFW Dresden, Helmholtzstr. 20, 01069 Dresden, Germany}
\author{D. V. Efremov} 
\affiliation{IFW Dresden, Helmholtzstr. 20, 01069 Dresden, Germany}
\author{J. M. Ok}
\affiliation{Department of Physics, Pohang University of Science and
Technology, Pohang 790-784, Korea}
\author{J. S. Kim}
\affiliation{Department of Physics, Pohang University of Science and
Technology, Pohang 790-784, Korea}
\author{Jeroen van den Brink} 
\affiliation{IFW Dresden, Helmholtzstr. 20, 01069 Dresden, Germany}
\affiliation{Department of Physics, Technische Universit\"at Dresden, 01062 Dresden, Germany}
\author{B. B\"uchner}
\affiliation{IFW Dresden, Helmholtzstr. 20, 01069 Dresden, Germany}
\affiliation{Department of Physics, Technische Universit\"at Dresden, 01062 Dresden, Germany}

\date{\today}

\begin{abstract}
{\bf
A very fundamental and unconventional characteristic of superconductivity in 
iron-based materials is that it occurs in the vicinity of two other 
instabilities.  
Apart from a tendency towards magnetic order,  these Fe-based systems have a 
propensity for nematic ordering: 
a lowering of the rotational symmetry while time-reversal invariance is preserved.
Setting the stage for superconductivity, it is heavily debated whether the 
nematic symmetry breaking is  
driven by
lattice, orbital or spin degrees of freedom.
Here we report a very clear splitting of NMR resonance lines in FeSe at 
$T_\text{nem}$ = 91K, far above the superconducting $T_c$ of 9.3 K.  
The splitting occurs for magnetic fields perpendicular to the Fe-planes and 
has the temperature dependence of a Landau-type order-parameter.  
Spin-lattice relaxation rates are not affected at $T_\text{nem}$, which 
unequivocally establishes orbital degrees of freedom as driving the nematic order. 
We demonstrate that superconductivity competes with the emerging nematicity.
}

\end{abstract}


\maketitle

Even if  the existence of  nematic order in the different classes of 
iron-based superconductors is by now a well-established experimental fact, its 
origin remains  
controversial~\cite{yi11,chu12,kasahara12,bohmer14,mcqueen09,Margadonna08,fu12}. 
It is related either to a lattice instability that causes a regular structural phase 
transition, to the formation of time-reversal invariant magnetic order, for instance a Ising 
spin-nematic~\cite{fang08a,xu08,fradkin10} state, or to the ordering of 
orbital degrees of freedom~\cite{kruger09,lv09,lee09b,daghofer10,chen10a}.  
As the nematic instability is a characteristic feature of the normal state 
from which at lower temperatures the superconductivity emerges, the different 
possible microscopic origins of nematicity are directly linked to the 
properties of  the superconducting state~\cite{fernandes14,paglione10}.  
From a symmetry point of view it is clear that when one of these three 
orderings (lattice/spin/orbital) develops, it must affect the other two -- the 
crucial challenge thus lies in establishing which ordering is primary, and to 
determine to which extend this primary order affects the two other degrees of 
freedom. 
It has been established that the lattice distortion, which at $T_\text{nem}$ 
reduces the crystallographic symmetry from tetragonal to orthorhombic, is an 
unlikely primary order parameter. Not only because the distortion is weak, but 
also because measurements of the resistance anisotropy have shown that the 
structural distortion is a conjugate field to a primary order parameter, 
therefore not the order parameter itself~\cite{chu12}. This basically 
restricts the driving force for the nematicity to be of electronic origin: 
either due the electron's spin or its orbital degree of freedom. 

\begin{figure}
\centering
\includegraphics[width=0.8\linewidth]{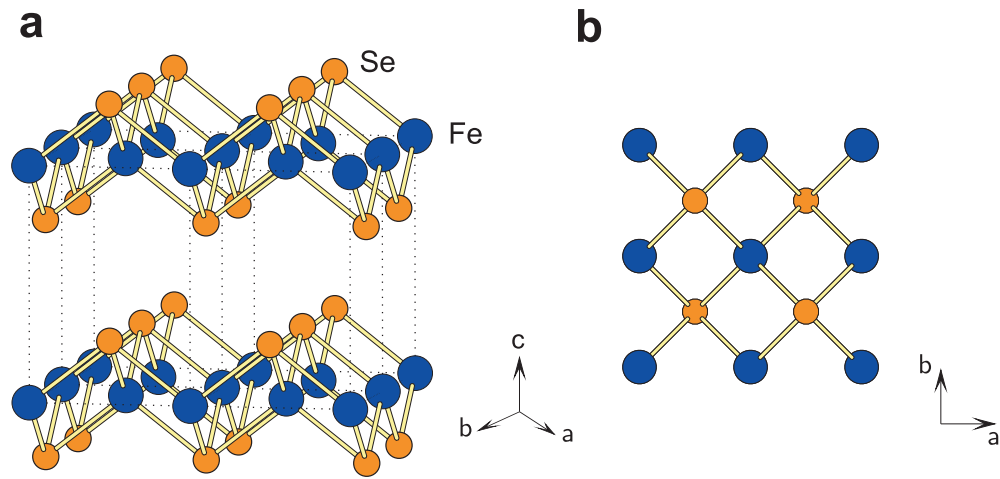}
\caption{{\bf Schematic crystallographic structure of FeSe. a,b,}  Fe-Se layers 
stacked along the $c$-direction (\bc{a}) and  the in-plane Fe atoms forming an 
almost square lattice with Se atoms centered alternately above and below Fe$_4$ 
plaquettes (\bc{b}). } 
\label{fig:1} 
\end{figure}

\fese\ is an attractive iron-based superconductor to study this issue, as 
it is a binary system with a rather simple structure (see Fig.~\ref{fig:1}), 
while sharing many common features with other Fe-based 
superconductors~\cite{hsu08}. Our bulk FeSe single-crystals undergo a clear 
tetragonal to orthorhombic transition at $T_\text{nem}$=91 K and at $T_c=9.3$ K 
superconductivity sets in, which is consistent with previous 
reports~\cite{mcqueen09}.  In single-layer FeSe films a much higher $T_c$  has 
been reported, 65 K~\cite{wang12b,xiang12,tan13,he13,zhang14} and 
above~\cite{ge14}, which is even higher than in any other iron-based 
superconductor. 
The high quality of our FeSe single crystals is confirmed by their very sharp 
superconducting transition and large residual resistivity ratio (see 
Supplementary Methods). 

\begin{figure}
\centering
\includegraphics[width=0.8\linewidth]{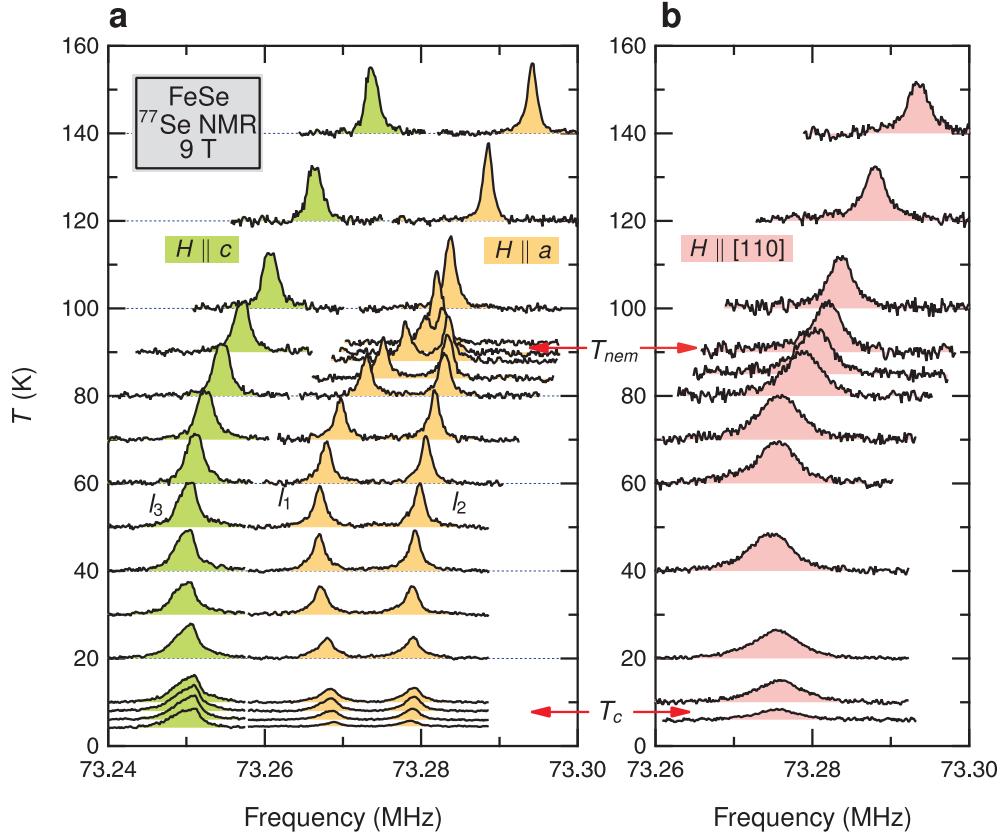}
\caption{{\bf \se\ NMR spectra for the FeSe single crystal.} \bc{a},  
	Measured at a field of 9 T applied parallel to either the 
	crystallographic $a$-axis or $c$-axis as a function of temperature.   
	The \se\ line splits into two lines ($l_1$ and $l_2$) at 
	$T_\text{nem}=91$ K for $H \parallel a$, while the line $l_3$ for $H\parallel c$ 
	remains narrow at all temperatures. To avoid an overlap, the spectra 
	for $H\parallel c$ are offset by $-10$ kHz. \bc{b}, 
	For an in-plane magnetic field where $H\parallel [110]$.   The absence 
	of the line splitting for this field orientation is direct proof for a 
	breaking of the local four-fold rotational symmetry.  The broadening 
	of the line is attributed to the strain induced by glueing  this 
	crystal inside the NMR coil.  
}
\label{fig:2} 
\end{figure}

To  establish whether spins or orbitals are responsible for its nematic 
instability we have measured \se\ NMR spectra as a function of temperature. 
The Se atoms in FeSe sit above and below at the center of the Fe$_4$ 
plaquettes tha form an almost square lattice (Fig.~\ref{fig:1}\bc{b}). 
For the NMR measurements we used an external field $H=9$T applied in a 
direction either parallel or perpendicular to the crystallographic $c$-axis, 
which is normal to the Fe planes (see Fig.~\ref{fig:1}\bc{a}). %
In the high-temperature tetragonal phase the spectra are extremely narrow with 
the full width at half maximum of $\sim$1 kHz for $H\parallel a$ and  $\sim$1.5 
kHz for $H\parallel c$, which is characteristic of a highly homogeneous 
sample (see Fig.~\ref{fig:2}). Below $T_\text{nem}$ we observe that the \se\ line 
splits into two lines with equal spectral weight for in-plane fields, 
$H\parallel a$. 
Note that in the orthorhombic phase our crystal is fully twinned.  The notation 
``$H\parallel a$'' thus means that actually one type of domains in the crystal 
experience a magnetic field $H\parallel a$ and the other type of domain has 
$H\parallel b$.  These two domains occur with equal probability.  
We shall refer to these lines as $l_1$ and $l_2$ with frequency 
$\nu_1$ and $\nu_2$, respectively ($\nu_1 < \nu_2$). In contrast, the \se\ 
spectrum for $H\parallel c$ consists of a single line $l_3$ at frequency 
$\nu_3$ that does not split and remains narrow down to low temperatures.  
From this, one can already conclude that the $l_1$-$l_2$ line splitting must be 
the consequence of an in-plane symmetry change.  

\begin{figure}
\centering
\includegraphics[width=1.0\linewidth]{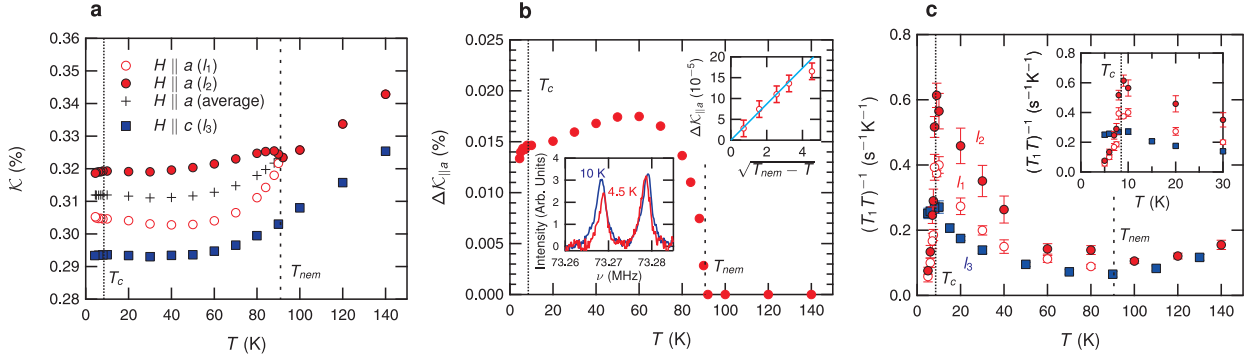}
\caption{{\bf Emergence of orbital-driven nematic state in FeSe.}
\bc{a}, Temperature dependence of the \se\ NMR Knight shift $\mathcal{K}$ for 
fields $\perp$ and $\parallel$ to the $c$-axis. Whereas at $T_\text{nem}$, 
$\mathcal{K_{\parallel a}}$ splits into lines $l_1$ and $l_2$, both 
$\mathcal{K}_{\parallel c}$ and $\mathcal{K}_{\parallel a}^\text{av}$, the average 
position of lines $l_1$ and $l_2$, show a smooth $T$-dependence. 
\bc{b}, Temperature dependence of the $l_1$-$l_2$ line splitting below 
$T_\text{nem}$ in terms of the difference in Knight shift 
$\Delta \mathcal{K}_{\parallel a}$. Inset: (upper right)  below $T_\text{nem}$ the splitting 
$\Delta \mathcal{K}_{\parallel a}$ is proportional to $\sqrt{T_\text{nem}-T}$, as is 
expected for an order parameter at a second order phase transition; (lower 
left) the comparison of two $^{77}$Se spectra at 10 K ($>T_c$) and 4.5 K ($<T_c$) 
reveals that the splitting between the $l_1$ and $l_2$ lines  clearly 
decreases in the superconducting state. The intensities of two spectra were 
normalized for comparison. 
\bc{c}, Temperature dependence of the spin-lattice relaxation rate divided by 
$T$, \slrt. The error bars reflect the uncertainty in the fitting procedure.  
At around $T_\text{nem}$ the spin relaxation rate barely changes,  
indicating the absence of a magnetic instability. Approaching $T_c$ enhances 
\slrt, signaling that AFM spin fluctuations develop. Below $T_c$,  as is 
conventional in the superconducting state, \slrt 
strongly drops.  The inset shows an enlargement of the low-temperature regime.
}
\label{fig:3} 
\end{figure}

We note that the \se\ nuclear spin is 1/2 so that the observed splitting 
cannot be due to a quadrupolar-type coupling to local lattice distortions.  
This is in contrast to LaFeAsO, in which the quadrupolar splitting of the  
$^{75}$As line in twinned  
single crystals for $H\perp c$  reflects the presence of orthorhombic 
domains~\cite{fu12}. 
On two further grounds it can be excluded that the orthorhombic lattice 
distortion causes the $l_1$-$l_2$ splitting. First, the splitting 
changes significantly when FeSe enters the superconducting state (see 
Fig.~\ref{fig:3}\bc{b}), where the lattice structure does not change 
notably~\cite{bohmer13}. That the splitting is of electronic origin is 
attested also by a more detailed consideration of the temperature dependence 
of the resonance frequency $\nu_i$ ($i=1...3$) for each of the three NMR 
lines. The $T$ dependence is shown in Fig.~\ref{fig:2} in terms of the Knight 
shift  $\mathcal{K}_i=(\nu_i-\nu_0)/\nu_0$ of $\nu_i$  away from an isolated 
nucleus ($\nu_0=\gamma_n H$ with the nuclear gyromagnetic ratio $\gamma_n$). 
In a paramagnetic state 
$\mathcal{K}=A_\text{hf}\chi_\text{spin}+\mathcal{K}_\text{chem}$ so that 
$\mathcal{K}$ is directly related to the local spin susceptibility \chispin. 
Here $A_\text{hf}$ is the hyperfine coupling constant and 
$\mathcal{K}_\text{chem}$ the temperature-independent chemical shift. 
It is clear that the splitting between $l_3$ and the degenerate $l_1$, $l_2$ 
pair in the tetragonal structure (that is, for $T>T_\text{nem}$)  is caused by the 
in-plane ($\parallel a$)--out-of-plane ($\parallel c$) anisotropy of the 
hyperfine coupling and the spin susceptibility. This anisotropy is caused by 
the crystallographic structure being very different in the directions $\parallel a$ 
and $\parallel c$, owing to the manifestly layered lattice structure of FeSe. 
From the data in Fig.~\ref{fig:3}\bc{a} it is clear that the $l_3$-$l_{1,2}$ 
splitting $\nu_3-\nu_{1,2}$ above $T_\text{nem}$ is similar in size to the $l_2$-
$l_1$ splitting $\Delta \nu = \nu_2-\nu_1$ in the low temperature orthorhombic 
state.  It is evident that such a very large splitting $\Delta \nu$ cannot be 
caused by the small lattice displacements in the orthorhombic state, involving 
atoms that move distances less than 0.5 \% of the lattice 
constant~\cite{mcqueen09,bohmer13}. This is exemplified by the average 
$\mathcal{K}_{\parallel a}^\text{av}=(\mathcal{K}_1+\mathcal{K}_2)/2$ of the two 
$H\parallel a$ and $H\parallel b$ lines (for the two different orthorhombic 
domains) having the same temperature dependence as  
$\mathcal{K}_{\parallel c}=\mathcal{K}_3$ in the entire temperature range.  
This is very different from the behaviour of the Knight shift splitting 
$\Delta \mathcal{K}_{\parallel a}=(\mathcal{K}_2-\mathcal{K}_1)/2 \propto \Delta \nu$ 
between $l_2$ and $l_1$ below $T_\text{nem}$. From the temperature dependence of 
$\Delta \mathcal{K}_{\parallel a}$ (shown in Fig.~\ref{fig:3}\bc{b}), one sees that it 
exhibits the typical $\sqrt{T_\text{nem}-T}$ behaviour of a Landau-type order 
parameter close to a second-order phase transition. 

Having established an order parameter type of behaviour of splitting 
$\Delta \nu$ and having excluded it is of lattice origin, we consider next the 
possibility that spin degrees of freedom cause the observed in-plane 
anisotropy of the Knight shift in the orthorhombic state.  
We have therefore measured the spin-lattice relaxation rate \slr\ as a 
function of temperature (see Fig. ~\ref{fig:3}). The quantity \slrt\ is 
proportional to the $\mathbf{q}$-sum of the imaginary part of the dynamical 
susceptibility, that is, 
$(T_1T)^{-1} \propto  
\sum_\mathbf{q} A_\text{hf}^2(\mathbf{q})\chi''(\mathbf{q},\omega)/\omega$, 
thereby probing antiferromagnetic (AFM) spin fluctuations.   
We observe that when crossing the nematic phase transition, \slrt\ barely 
changes, indicating that AFM fluctuations are not enhanced around $T_\text{nem}$ 
and the system is evidently very far away from any magnetic instability. Only 
when further lowering the temperature we observe that  \slrt\ gradually 
increases and that at $T_c$, when superconductivity sets in, the AFM 
fluctuations are significantly enhanced.  
This observation is in agreement with previous \slrt\ measurements on FeSe 
powders~\cite{imai09} and evidences that spin fluctuations are not driving the 
nematic transition. Moreover, the extremely narrow \se\ NMR lines being well 
preserved down to 4.2 K, indicates the complete absence of static magnetism 
\cite{medvedev09}. 

\begin{figure}
\centering
\includegraphics[width=0.7\linewidth]{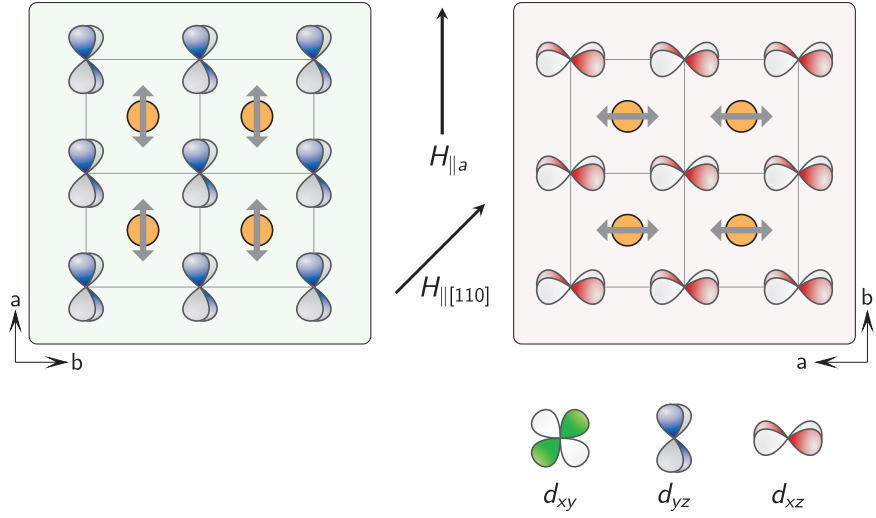}
\caption{{\bf Top view of the FOO in FeSe with the two 
different domains that are present in a twinned crystal.} The three orthogonal 
orbitals $d_{xy}$,  $d_{yz}$ and  $d_{xz}$ are indicated. The  double-headed 
arrow indicates the nematic order parameter. } 
\label{fig:4}
\end{figure}

The remaining degree of freedom that can drive the nematic ordering is the 
orbital one, in particular in the form of ferro-orbital order (FOO). It is 
clear that such an orbital ordering breaks the in-plane local symmetry at the 
Se sites (see Fig.~\ref{fig:4}), and generates two non-equivalent directions 
$\perp c$: the $a$ and $b$ direction. 
We first consider FOO from a theoretical point of view, defining the FOO order 
parameter as $\psi = (n_x-n_y)/(n_x+n_y)$, where $n_{x,y}$ corresponds to the 
occupation of $x = d_{xz}$ and $y = d_{yz}$ orbitals indicated in the 
Fig.~\ref{fig:4} ($z$ corresponds to the crystallographic $c$-axis). Given the 
symmetries of the system, the free energy in the vicinity of the orbital 
ordering transition in the presence of a magnetic field $\mathbf{H}$ can be 
expanded as: 
\begin{equation}
F = \frac{a}{2} \psi^2 + \frac{b}{4} \psi^4 + \frac{1}{2\chi_\perp} M^2  -  \gamma \psi (M_x^2 - M_y^2) +  \frac{g}{2} M_z^2+\mathbf{M H},
\label{eq:F}
\end{equation}
where $\mathbf{M}$ is the magnetic moment. An important quantity is $\gamma$, 
the coupling between the orbital order parameter and magnetization. For 
localized $3d$ states it is perturbatively related to 
the strength of spin-orbit interaction $\lambda$ and energy difference 
$\Delta_d$ between the $x$ and $y$, $z$ states as $\gamma\propto \lambda^2/\Delta_d$.
From Eq.~\ref{eq:F} one obtains susceptibilities of the form 
$\chi_{xx,yy} = \chi_\perp/(1 \pm \gamma \psi \chi_\perp ) \approx 
\chi_\perp(1 \mp \gamma \psi \chi_\perp ) $
and $\chi_{zz} = \chi_\perp/(1+\chi_\perp  g)$. Owing to the linear coupling the 
orbital order parameter is directly proportional to the anisotropy in the 
magnetic susceptibility: 
$\chi_{xx}-\chi_{yy} \propto \psi \propto \sqrt{T_{OO} - T}$ in the vicinity 
of the ferro-orbital ordering transition.  

Now the question arises how such a ferro-orbital ordering affects the Knight 
shifts 
$\mathcal{K}_{\alpha} = A^\text{hf}_{\alpha \alpha} \chi_{\alpha \alpha}$, 
where $\alpha = x,y,z$.  Owing to the orthorhombic symmetry only the three 
diagonal terms are present~\cite{chu12}. 
This is in agreement with the experiments showing 
$\mathcal{K}_{x,y} \neq \mathcal{K}_{z}$. It is  useful to consider the 
average, isotropic part of the in-plane Knight shift 
$ \mathcal{K}_{\parallel a}^{av} = A^\text{hf}_{xx} \chi_{xx} + A^\text{hf}_{yy} \chi_{yy} = 1/2(A^\text{hf}_{xx}  + A^\text{hf}_{yy}) ( \chi_{xx}+\chi_{yy})+1/2(A^\text{hf}_{xx}  - A^\text{hf}_{yy}) ( \chi_{xx}-\chi_{yy})$  
separately from the difference, the anisotropic in-plane Knight shift
$\Delta \mathcal{K}_{\parallel a} = A^\text{hf}_{xx} \chi_{xx} - A^\text{hf}_{yy} \chi_{yy} = 1/2(A^\text{hf}_{xx}  + A^\text{hf}_{yy})( \chi_{xx}-\chi_{yy})+1/2(A^\text{hf}_{xx}  - A^\text{hf}_{yy})( \chi_{xx}+\chi_{yy})$.
An analysis of the hyperfine constants establishes that 
$A^\text{hf}_{xx}  - A^\text{hf}_{yy} \propto \psi$, so that 
$\Delta\mathcal{K}_{\parallel a} \propto \psi \propto \sqrt{T_{OO} - T}$. Thus the 
anisotropic Knight shift is directly proportional to the orbital ordering 
parameter but the same analysis shows that $\mathcal{K}_{\parallel a}^{av}$ 
and $\mathcal{K}_{z}$ may  
depend on $\psi$ only in higher order.  

We can now compare the theoretical analysis for an orbital-driven nematic 
state with our experimental results. Clearly the measured splitting 
$\Delta \mathcal{K}_{\parallel a}$ shows the $\sqrt{T_\text{nem} - T}$ behaviour close to 
the critical temperature, so that we conclude that $T_\text{nem} = T_{OO}$. At the 
same time the measured $\mathcal{K}_{\parallel a}^{av}$ and 
$\mathcal{K}_{\parallel c}$ 
(see Fig.~\ref{fig:3}\bc{a}) indeed barely show an anomaly in their 
temperature dependence.  
In the normal state, between $\sim 50-60$K and $T_c$ the splitting 
$\Delta \mathcal{K}_{\parallel a}$ decreases. This is due to the two 
distinct contributions to $\Delta \mathcal{K}_{\parallel a}$: the temperature 
dependence of the hyperfine constant $A^\text{hf}_{xx} - A^\text{hf}_{yy}$ and 
of the susceptibility $\chi_{xx}-\chi_{yy}$. The former saturates below $50-60$K, 
as the nematic order parameter  tends to a constant~\cite{bohmer13}. At the 
same time the anisotropic part of the transverse susceptibility changes owing to 
non-Fermi liquid effects caused by the enhanced spin 
fluctuations~\cite{Korshunov09}, leading to the observed decrease in 
$\Delta \mathcal{K}_{\parallel a}$ in the normal state.  
The issue that remains open from the NMR data is the precise pattern of 
orbital ordering  that is formed. The NMR data does not fix the directions of 
$x$ and $y$ with respect to the crystallographic axes. Any rotation of the FOO 
orbital ordering pattern around the $c$-axis is therefore possible in 
principle. However, the orthorhombic lattice distortion induced by the FOO 
ordering leaves all Fe-Se distances equivalent~\cite{Margadonna08}, which 
implies that $x \parallel a$ and $y \parallel b$, leading to the FOO pattern 
in Fig.~\ref{fig:4}. 

The conclusion above, that below $T_\text{nem}$ the orbital order as shown in 
Fig.~\ref{fig:4} renders the electronic structure along the $a$ and $b$ 
direction inequivalent, resulting in a clearly different NMR responses for 
$H\parallel a$ and $H\parallel b$, can be tested. When the magnetic field is 
applied in the $ab$ plane in the diagonal direction, that is. 
$H\parallel [110]$, the field has equal projections on $a$ and $b$ (see Fig. 4).
Therefore, the two domains in our twinned crystal should now yield the 
same NMR response, implying that for this field orientation the 
splitting between $l_1$ and $l_2$ in the orthorhombic state below $T_\text{nem}$ 
should be absent.   We performed the experiment with $H\parallel [110]$, using 
a different single-crystalline platelet glued in the required orientation.  As 
shown in Fig. 2\bc{b} now a splitting of the line below $T_\text{nem}$ is indeed 
clearly absent, which is direct proof that below $T_\text{nem}$ the rotational ($C_4$) 
symmetry is broken.  
We note that our NMR experiments do not provide information on the size of the 
domains, which might in principle be ordered or disordered at a microscopic 
scale, which implies the presence of a certain amount of antiferro orbital 
ordering. The relevance of such secondary orderings might be probed by NMR 
experiments on detwinned crystals. 

Having established that the orbital order drives the nematic ordering, 
the question arises how the orbital ordering affects not only the 
lattice and spin degrees of freedom, but also the superconducting state. The 
relation to the secondary orthorhombic lattice distortion has been discussed 
above. From the NMR data also the coupling between the orbital order to the 
spin degrees of freedom is directly evident. 
The spin-lattice relaxation rate, measuring the strength of low-energy spin fluctuations, 
shows that in the vicinity of $T_\text{nem}$ there is little, if any, enhancement 
of the magnetic excitations, an enhancement that would be expected from 
Fermi-liquid theory in the vicinity of a spin-density wave transition.   This implies that 
the characteristic energy of the degrees of freedom driving the nematic 
transition considerably differs from the characteristic energy of magnetic 
degrees of freedom: orbital and spin degrees of freedom are well separated.  
When going below $T_\text{nem}$ the spin-lattice relaxation rate increases 
steadily,  approaching $T_c$ in a 
manner that is quantitatively different for the lines $l_1$ and $l_2$ (see 
Fig.~\ref{fig:3}\bc{c}). This is to be expected because the spin-relaxation rate 
in the FOO state picks up the anisotropies in its hyperfine couplings and 
susceptibilities, as the Knight shift does. 

Finally, we analyse the interplay of the orbital ordering and 
superconductivity. Previously scanning tunneling spectroscopy measurements 
have found a two-fold breaking of the Cooper-pair symmetry in FeSe, which 
implies that the superconducting order parameter is directly affected by the 
nematicity~\cite{song11}.  
Here we observe the complementary effect: the splitting 
$\Delta \mathcal{K}_{\parallel a}$ (which is proportional to the orbital order 
parameter) changes significantly below $T_c$ (see Supplementary Methods). 
Thus, the nematic order parameter  
is directly affected by superconductivity. The splitting 
$\Delta \mathcal{K}_{\parallel a}$ becoming  smaller while the Knight shifts 
$\mathcal{K}_{\parallel c}$ and $\mathcal{K}^{av}_{\parallel a}$ barely change indicates 
that in our bulk FeSe crystals superconductivity and nematicity 
compete---superconductivity tends to suppress orbital ordering and vice versa. 
It is  
interesting to note that in the single layers of FeSe for which spectacularly 
high $T_c$ values have been 
reported~\cite{wang12b,xiang12,tan13,he13,zhang14,ge14} a 
tetragonal-orthorhombic transition is absent, evidencing a much weaker nematic 
tendency. It will have to be established whether the suppressed nematicity is 
a cause for the strongly enhanced $T_c$ in FeSe single layers. 


\subsection*{Methods}

Single crystals of \fese\ ($T_c\sim 9.3$ K) were grown using a KCl-AlCl$_3$ flux techniques as described in detail elsewhere \cite{chareev13}. The mixture of Fe, Se, AlCl$_3$ and KCl were
sealed in evacuated pyrex ampoule. The samples were heated to 450 $^\circ$C in 
a horizontal tube furnace, held at this temperature for 40 days. The 
temperature of the hottest part of the ampoule was 450 $^\circ$C and the 
coolest part was 370--380 $^\circ$C. The obtained product was washed with 
distilled water to remove flux and other by-products and then the 
tetragonal-shaped single crystals were mechanically extracted. The typical 
size of obtained crystal was $1 \times 1 \times 0.1$ mm$^3$. The temperature dependence of 
resistivity of \fese\ single crystals was measured using conventional 
four-probe configuration in a 14T physical property measurements system(PPMS) 
and the magnetic susceptibility was measured in 5T magnetic property 
measurements system (MPMS). 

$^{77}$Se (nuclear spin $I=1/2$) NMR was carried out in a \fese\ single 
crystal ($0.7\times0.7\times0.1$ mm$^3$)  at an external field of 9 T and in 
the range of temperature 4.2 --- 140 K. The sample was rotated using a 
goniometer for the exact alignment along the external field. The \se\ NMR 
spectra were acquired by a standard spin-echo technique with a typical $\pi/2$ 
pulse length 2--3 $\mu$s. 
The nuclear spin-lattice relaxation rate \slr\ was obtained by fitting the 
recovery of the nuclear magnetization $M(t)$ after a saturating pulse to a 
single exponential function, $1-M(t)/M(\infty)=A \exp(-t/T_1)$ where $A$ is a 
fitting parameter. 


\bibliography{mybib}



\let\origdescription\description
\renewenvironment{description}{
  \setlength{\leftmargini}{0em}
  \origdescription
  \setlength{\itemindent}{0em}
  \setlength{\labelsep}{\textwidth}
}
{\endlist}

\begin{description}
\item[Acknowledgment] The authors thank G. Prando and H.-J. Grafe for 
	discussion. This work has been supported
 	by the Deutsche Forschungsgemeinschaft (Germany) via DFG Research 
	Grants BA 4927/1-1 and the Priority Program SPP 1458. Financial 
	support through the DFG Research Training Group GRK 1621 is gratefully 
	acknowledged. The work at POSTECH was supported by 
 	the National Research Foundation (NRF) through the Mid-Career Researcher Program
	(No. 2012-013838), SRC Center for Topological Matter (No. 2011-0030046), and the Max
	Planck POSTECH/KOREA Research Initiative Program (No. 2011-0031558), and also by
	Institute of Basic Science (IBS) through Center for Artificial Low 
	Dimensional Electronic Systems.
\item[Author Contributions] SHB performed the main NMR measurements, analyzed 
	data, and participated in writing of the manuscript; JO and JK 
	synthesized the sample; DE and JvdB provided theoretical support and 
	participated in writing of the manuscript; BB supervised and guided 
	the study and participated in the writing of the manuscript. All 
	authors discussed the results and commented on the manuscript. 
\item[Competing financial interests] The authors declare no 
	competing financial interests. 
\item[Additional information] Correspondence and requests for materials should be 
	addressed to S.-H. Baek~(email: sbaek.fu@gmail.com). 

\end{description}

\pagebreak

\clearpage

\setcounter{figure}{0}
\renewcommand{\thefigure}{S\arabic{figure}}

{\bf Supplementary Material to ``Orbital-driven 
nematicity in FeSe''}

\subsection{Sample characterization}

The stoichiometry of Fe and Se is confirmed by energy dispersive X-ray 
spectroscopy (EDS) on the cleaved  
surface of a single crystal. The single crystal X-ray diffraction (XRD) 
shows only (001) peaks,  
as shown in Fig. S1\bc{a}, revealing the good crystallinity 
of single tetragonal phase crystals. 

\begin{figure}[b!]
\centering
\includegraphics[width=\linewidth]{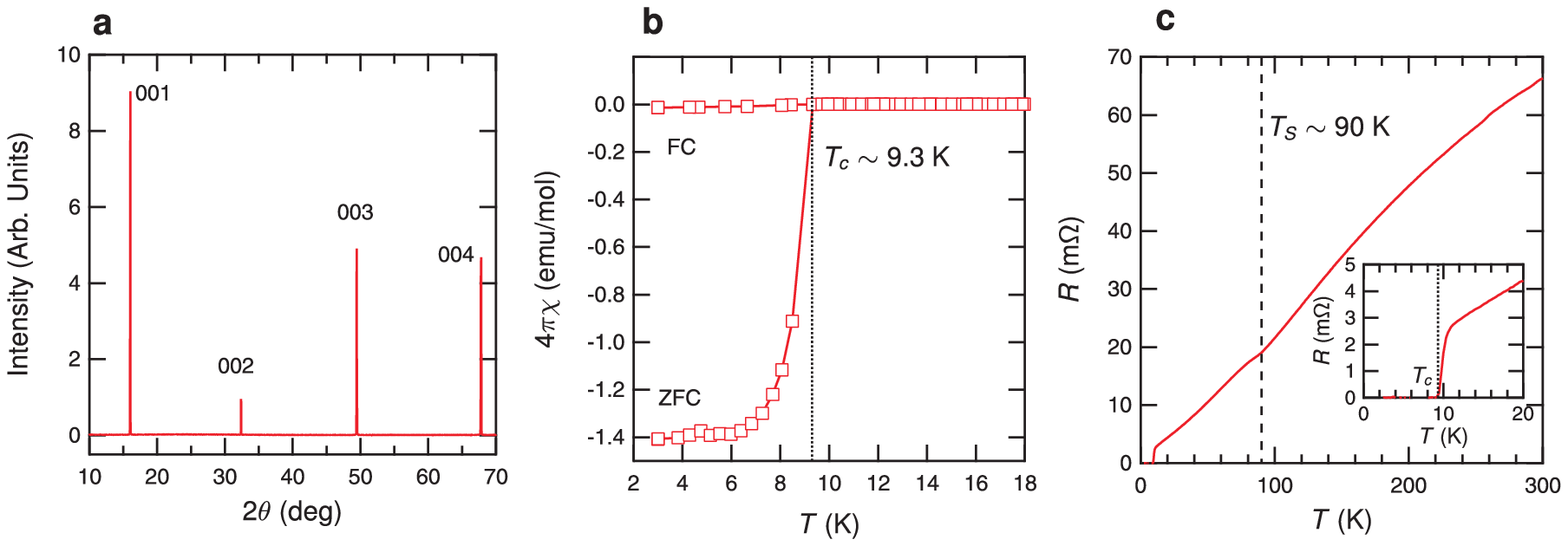}
\caption{ (\bc{a}) Single crystal XRD pattern shows only the  (00N) reflections of the tetragonal FeSe phase.  (\bc{b})  Uniform magnetic susceptibility $\chi$ and (\bc{c}) in-plane resistance $R$ as a function of temperature. }
\end{figure}

Figure S1\bc{b} shows the uniform magnetic susceptibility $\chi$ as a function of 
temperature in a FeSe single crystal measured at $H = 10$ Oe. $\chi(T)$ 
shows a very sharp superconducting transition at $T_c\sim9.3$ K, which is  
higher than that reported in literature so far \citeS{hu11,bohmer13b}.
Figure S1\bc{c} presents the temperature dependence of resistance $R$ 
under zero magnetic field. We find a small kink at $\sim90$ K due  
to the structural phase transition from tetragonal to orthorhombic phase 
\citeS{mcqueen09b,hu11}. The 
residual resistivity ratio (RRR) of the sample was found to be RRR 
= $R$(300 K)/$R$(11 K) $\sim 30$ which is much larger than previous results 
\citeS{hu11,vedeneev13}. $R(T)$ also shows a sharp 
superconducting transition with the onset $T_c^\text{onset}\sim 10.5$ K.   
Note that $R(T)$ becomes zero at $T_c$ determined from $\chi(T)$.

Therefore, the very sharp superconducting transition with $T_c\sim 9.3$ K 
as well as the very large RRR confirm the high quality of our FeSe single 
crystals.

\subsection{NMR in the superconducting state}

The kine-shifts of $l_1$ and $l_2$ in the superconducting state are shown in 
detail in Figure S2\bc{a}. The lines shift symmetrically toward each other as 
temperature is lowered with a change of   
$\Delta\nu$ in the SC state of about 10\% at 4.2 K $\sim  T_c/2$.  The spin 
relaxation rate in the SC state is shown in Figure S2\bc{a}. It is clear that 
\slr\ of both $l_1$ and $l_2$ probes the   
same $T_c\sim 8.5$ K and in the temperature range  between $T_c /2$ and $T_c $ 
shows a power law dependence $T_1^{-1} \propto T^5$, which is significantly 
steeper than the $T^3$ dependence  
reported in Ref. \citeS{kotegawa08}.   

\begin{figure}
\centering
\includegraphics[width=0.8\linewidth]{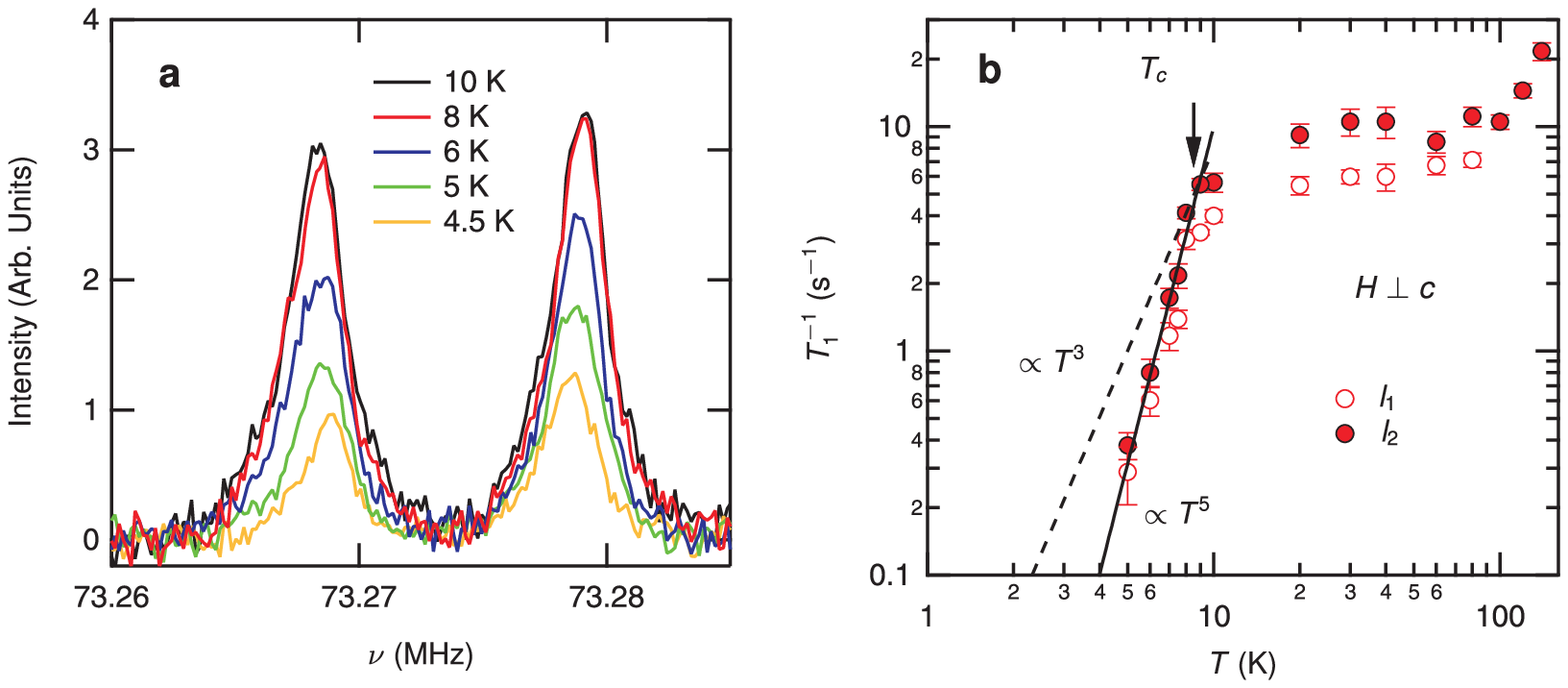}
\caption{(\bc{a}) Temperature evolution of \se\  spectra in the SC state for $H \perp c$.  $T$ was multiplied to each spectrum  for a Boltzmann correction.   With decreasing $T$, both $l_1$ and  $l_2$ peaks  rapidly lose their intensities due to the superconducting shielding effects and,  at the same time, shift toward each other.  (\bc{b}) A log-log plot of \slr\ vs. $T$ shows the power  
law behavior $T_1^{-1}\propto T^5$ for both $l_1$ and $l_2$ in the SC state.}
\end{figure}

\bibliographystyleS{mbplane}

\end{document}